\documentclass[%
 aip,
 amsmath,amssymb,
 reprint,%
 fleqn,
]{revtex4-1}

\usepackage{graphicx}% Include figure files
\usepackage{dcolumn}% Align table columns on decimal point
\usepackage{bm}% bold math
\usepackage[outdir=./]{epstopdf} 
\usepackage{amsfonts} 
\usepackage[utf8]{inputenc}
\usepackage[T1]{fontenc}
\usepackage{mathptmx}
\begin{document}

\preprint{AIP/123-QED}

\title{Adaptation rules inducing synchronization of heterogeneous Kuramoto oscillator network with triadic couplings}
% Force line breaks with \\

\author{Anastasiia A. Emelianova}
\email{emelianova@ipfran.ru}
\affiliation{ 
A.V. Gaponov-Grekhov Institute of Applied Physics of the Russian Academy of Sciences, 46 Ulyanov Street, 603950, Nizhny Novgorod, Russia %\\This line break forced with \textbackslash\textbackslash
}%

\author{Vladimir I. Nekorkin}
\email{vnekorkin@appl.sci-nnov.ru}
\affiliation{ 
A.V. Gaponov-Grekhov Institute of Applied Physics of the Russian Academy of Sciences, 46 Ulyanov Street, 603950, Nizhny Novgorod, Russia %\\This line break forced with \textbackslash\textbackslash
}%
\affiliation{ 
National Research Lobachevsky State University of Nizhny Novgorod, 23 Gagarin Avenue, 603022, Nizhny Novgorod, Russia %\\This line break forced with \textbackslash\textbackslash
}%

\date{\today}% It is always \today, but any date may be explicitly specified

\begin{abstract}

A class of adaptation functions is found for which a synchronous oscillation mode exists in the network of phase oscillators with triadic couplings. It is shown that the destruction of the synchronous mode occurs differently for networks with pairwise couplings and with higher-order interactions. In the first case, a chimera state is realized. In the second case, the destruction of the synchronous state occurs more abruptly, and the chimera state is not formed. The patterns of formation of synchronization and desynchronization modes are determined.

\end{abstract}

\maketitle

\begin{quotation}

An important problem in nonlinear science is the study of complex networks. At the same time, in many real networks dynamic processes occur simultaneously in nodes and in couplings between elements. An example is the neural networks of the brain, where one need to consider simultaneously the electrochemical activity of individual cells and the activity of axonal connections between them. Such systems are well described by the concept of adaptive networks. Herewith, the elements can be connected not in pairs, but in more complex geometric shapes. In this work, we study synchronization and desynchronization modes in a network of inhomogeneous oscillators with adaptive triadic couplings and study the dependence of synchronous states on the adaptation rule.

\end{quotation}

\section{\label{introduction} Introduction}
Nowadays, an important problem in nonlinear science is the study of complex networks. In this regard, the most difficult task is to establish the structural properties of the network and to answer the question of what principles are used to form a particular topology and what laws govern the evolution of the network. It is important to take into account that in many systems dynamic processes occur simultaneously in nodes and in couplings between elements. An example is the neural networks of the brain, where it is necessary to consider simultaneously the electrochemical activity of individual cells and the activity of axonal connections between them. In this regard, the concept of adaptive dynamic networks emerged. \cite{masl_adaptive,berner_adaptive} Using models of adaptive couplings, it is possible to describe, for example, synaptic plasticity (spike timing-dependent plasticity) \cite{adaptive_STDP}, chemical \cite{adaptive_chem}, physiological \cite{adaptive_phys}, social \cite{adaptive_social} systems, power grids \cite{adaptive_power}.

Elements of adaptive networks are usually connected in pairs. However, recent studies show that in many physical systems, there are also group-level interactions. Therefore, some systems can be better described in terms of higher-order interactions, when the networks elements form a geometric objects of different dimensions which reflect the underlying geometry of complex systems \cite{boccaletti2023}. Such properties are observed, for example, in functional \cite{func_brain1,func_brain2, func_brain3} and structural \cite{struct_brain} brain networks, protein interaction networks \cite{protein}, semantic networks \cite{semantic} and co-Authorship graphs in science \cite{collab}. 

Until recently, investigations of synchronization have been done either in static simplicial complexes or in temporal networks where group interactions are not taken into consideration. In the work [\onlinecite{temp_simp_comp}], an approach to the study of synchronization in temporal simplicial complexes is given for the general form of the right side of differential equations. However, this approach assumes that the element interaction function is set to zero in the synchronous state, which is not always true. The article [\onlinecite{kach_jal}] presents an approach to studying synchronization in adaptive simplicial complexes with initial conditions in the form of a delta function, which does not imply the zeroing of the interaction function. In this paper, we use this approach applied to a network of Kuramoto oscillators with adaptive couplings and study the influence of the adaptation rule on synchronization in the system, as well as the transient processes accompanying the process of loss of synchronization.

The paper is organized as follows. The details of the model are introduced in Sec. \ref{sec:model}. In Sec. \ref{sec:stationary_states}, we describe the stationary states of the network. In Sec. \ref{sec:mean-field}, we use the mean-field approach. In Sec. \ref{sec:numerical}, we demonstrate results of our numerical simulation. In Sec. \ref{sec:loss}, we describe the process of loss of stability of the synchronous state. In Sec. \ref{sec:conclusion}, we provide a summary of our main results.

\section{\label{sec:model} The model}

Consider a network of all-to-all Kuramoto oscillators with adaptive triadic couplings:
\begin{eqnarray}
&& \frac {d\theta_i}{dt} = \omega_i + \frac {\Lambda} {N^2} \sum_{j,k=1}^N \kappa_{ijk} \sin(\theta_j + \theta_k - 2\theta_i), \notag \\ 
&& \frac {d \kappa_{ijk}}{dt} = -\varepsilon \Bigl( \sin(\theta_j + \theta_k - 2\theta_i + \beta) + \kappa_{ijk} \Bigr). \label{sys_3D}
\end{eqnarray}

Here, $\theta_i \in (-\pi, \pi]$ are the phases of the oscillators, and $\kappa_{ijk}$ are the coupling weights. The parameter $\Lambda$ is the coupling strength, $\varepsilon$ separates the speed of the dynamics of the phases and of the dynamics of adaptation, and the parameter $\beta$ ($0 < \beta < 2\pi$) allows one to switch between the different adaptation rules. Natural frequencies are subject to the uniform distribution $\omega_i \sim U(-\Delta, \Delta)$. The initial phases are chosen to be 0 with probability $\eta$ and equal to $\pi$ with probability $(1-\eta)$. If we consider each oscillator as a neuron, we suppose that the phase is equivalent to the membrane potential. Then we can assume that the neuron generates a spike at the moment when the phase crosses $\theta_i=\pi$, and the neuron is at rest at $\theta_i=0$. This means that we take neurons at the initial moment of time either in a state of rest or in a state of spiking. The initial weights are $\kappa_{ijk}(0) = 1$. The number of elements is chosen $N=100$. Next in the paper, we fix the parameters $\Lambda = 4$ (except for FIG. \ref{fig:beta_K_R2}), $\varepsilon=1$ and $\Delta=1$.  

Below we compare the results obtained for a system with triadic connections with the results for pairwise connections. Therefore, for comparison, we consider a system of adaptively coupled oscillators with pairwise couplings in the following form:
\begin{eqnarray}
&& \frac {d\theta_i}{dt} = \omega_i + \frac {\Lambda} {N} \sum_{j=1}^N \kappa_{ij} \sin(\theta_j - \theta_i), \notag \\ 
&& \frac {d \kappa_{ij}}{dt} = -\varepsilon \Bigl( \sin(\theta_j - \theta_i + \beta) + \kappa_{ij} \Bigr). \label{sys_2D}
\end{eqnarray}

The system \eqref{sys_2D} was studied in [\onlinecite{kas2017}] in the case of homogeneous ($\omega_i = 1$) oscillators with time delay. It was shown that the system demonstrates self-organized emergence of hierarchical multilayered structures and chimera states. In [\onlinecite{kach_jal}], the system \eqref{sys_3D} was studied in the case $\beta = \frac {3\pi}{2}$, which corresponds to the Hebbian rule of plasticity\cite{hebb}. It was demonstrated that with triadic adaptive couplings, an abrupt transition to desynchronization occurs when the coupling strength parameter $\Lambda$ varies.

\section{\label{sec:stationary_states} Stationary states of the network}

First of all, we consider stationary states of the system \eqref{sys_3D}. 

In a stationary state, $\frac {d \kappa_{ijk}}{dt} \approx 0$, so from the second equation of \eqref{sys_3D} we get $\kappa_{ijk} \approx -\sin(\theta_j + \theta_k - 2\theta_i + \beta)$. It means that $\kappa_{ijk} \approx -1$ if $\theta_j + \theta_k - 2\theta_i + \beta \approx \frac {\pi}{2}$ and $\kappa_{ijk} \approx 1$ if $\theta_j + \theta_k - 2\theta_i + \beta \approx \frac {3\pi}{2}$. Therefore, eliminating the factor $\kappa_{ijk}$ in the first equation of the system \eqref{sys_3D}, we get 
\begin{equation}
\frac {d\theta_i}{dt} = \omega_i - \frac {\Lambda}{2} \cos\beta + \frac {\Lambda}{2 N^2} \text{Re} \Bigl \{ e^{i(\beta-4\theta_i)} \sum_{j,k=1}^N e^{2i(\theta_j + \theta_k)} \Bigr \}, \label{theta_eq1}
\end{equation}
or, taking into account that $R_2 e^{i\Psi_2} = \frac 1 N \sum_{j=1}^N e^{2 i \theta_j}$,
\begin{equation}
\frac {d\theta_i}{dt} = \omega_i - \frac {\Lambda}{2} \cos\beta + \frac {\Lambda R_2^2} {2} \cos(2 \Psi_2 - 4\theta_i + \beta). \label{theta_eq2}
\end{equation}

Transitioning to a reference frame rotating with a common frequency, we obtain $\Psi_2 = 0$. Then the synchronous states are given by the equation 
\begin{equation}
\cos(4\theta_i - \beta) = -\frac {2} {\Lambda R_2^2} ( \omega_i - \frac {\Lambda}{2} \cos\beta ). \label{theta_eq3}
\end{equation}

If $|\Gamma| < 1$, where $\Gamma = -\frac {2} {\Lambda R_2^2} ( \omega_i - \frac {\Lambda}{2} \cos\beta )$, then each oscillator has two stationary states
\begin{eqnarray}
&& \theta_{i}^{1*} = \frac {\beta + \text{arccos} \Gamma }{4}, \quad \theta_{i}^{2*} = \frac {\beta - \text{arccos} \Gamma }{4}. \label{theta_stat}
\end{eqnarray}

The characteristic exponents for them are
\begin{eqnarray}
&& \lambda(\theta_i^{1*}) = -2 \Lambda R_2^2 \sqrt{ 1 - \Gamma^2 } < 0, \notag \\
&& \lambda(\theta_i^{2*}) = 2 \Lambda R_2^2 \sqrt{ 1 - \Gamma^2 } > 0, \label{lyap_theta_stat}
\end{eqnarray}
so $\theta_i^{1*}$ is the stable state and $\theta_i^{2*}$ is the unstable one. As the parameters vary, these states disappear simultaneously through a saddle-node bifurcation.

\section{\label{sec:mean-field} Mean-field approach}

\begin{figure*}
\begin{minipage}[h]{0.47\linewidth}
\center{\includegraphics[width=1.\linewidth]{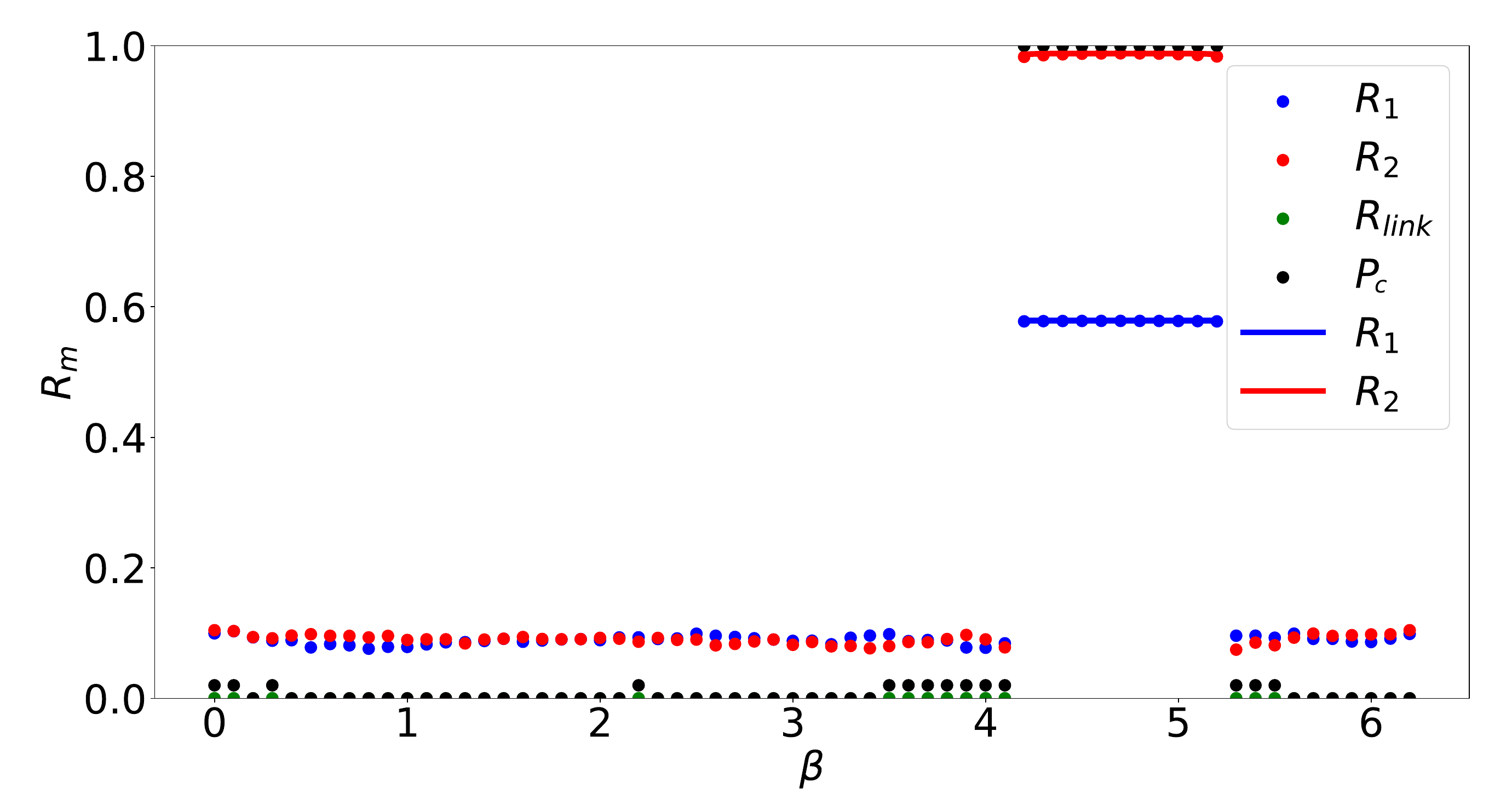}} \\ (a)
\end{minipage}
\hfill
\begin{minipage}[h]{0.47\linewidth}
\center{\includegraphics[width=1.1\linewidth]{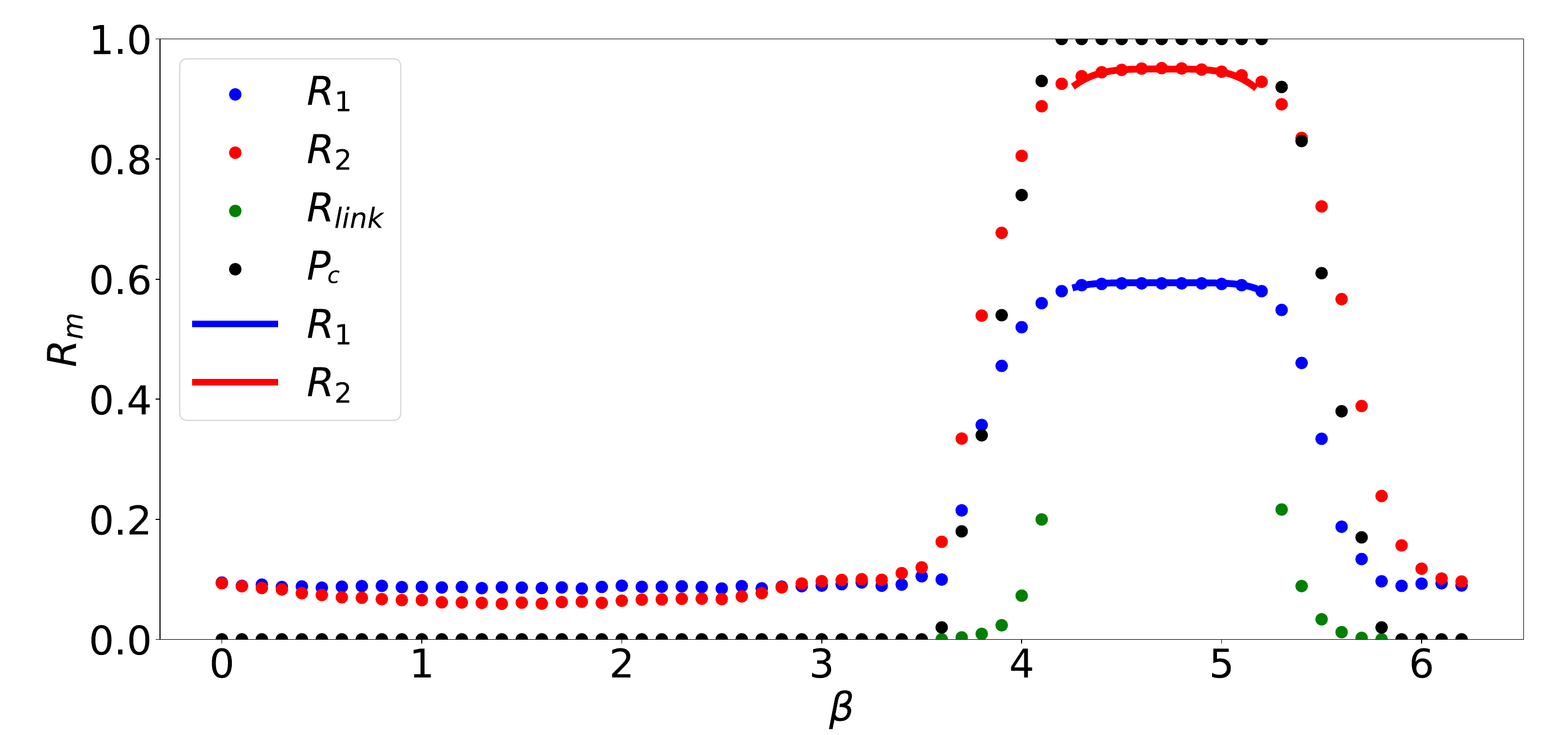}} \\ (b)
\end{minipage}
\caption{\label{fig:beta_Rm} Order parameters $R_1$, $R_2$ and $R_{link}$, $P_c$ depending on the parameter $\beta$ for $\eta=0.8$. Theoretical values are shown by solid lines, the results of numerical simulation are shown by dots. (a) Triadic couplings. (b) Pairwise couplings. }
\end{figure*}

\begin{figure*}
\begin{minipage}[h]{0.47\linewidth}
\center{\includegraphics[width=1.1\linewidth]{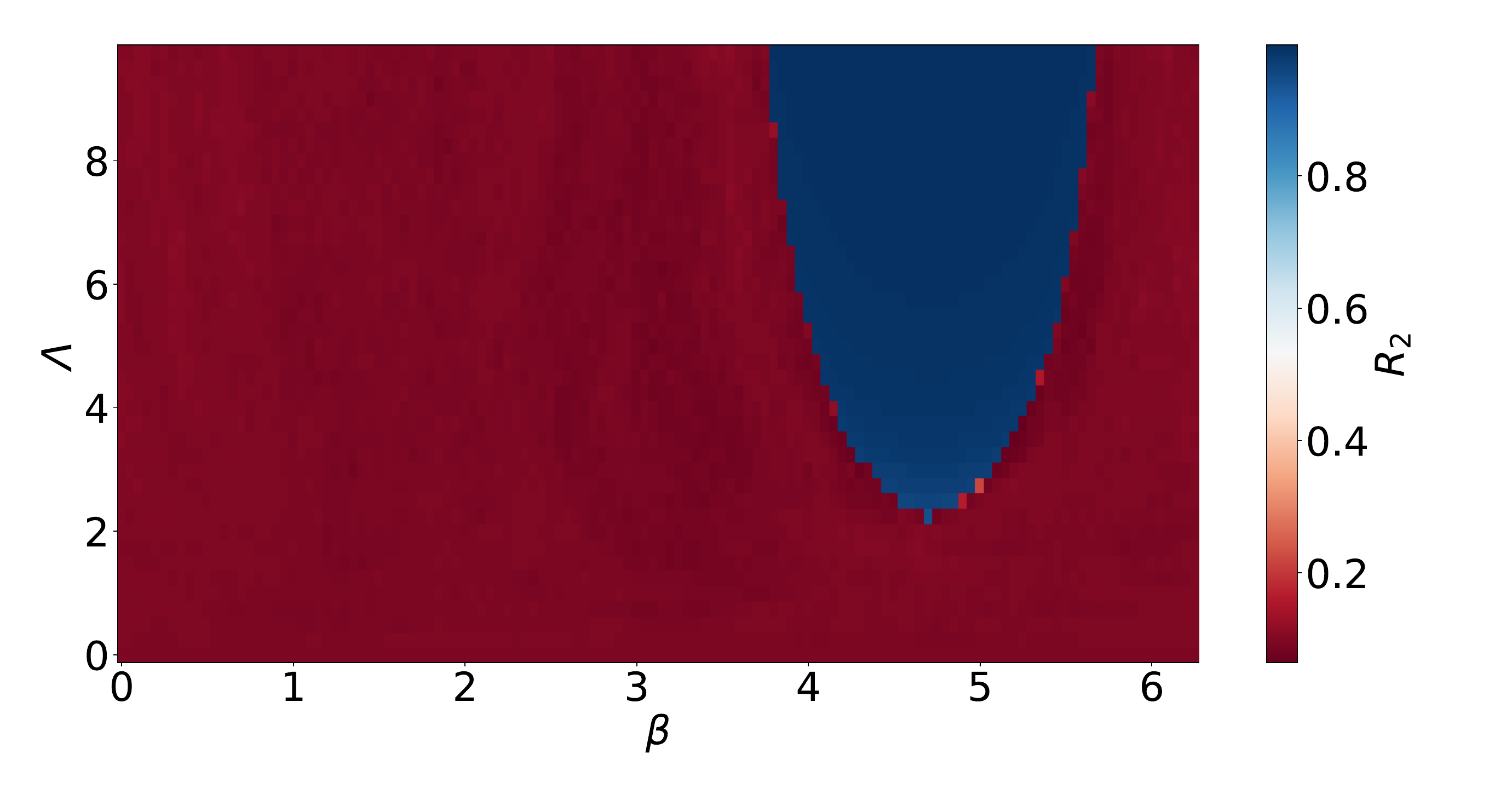}} \\ (a)
\end{minipage}
\hfill
\begin{minipage}[h]{0.47\linewidth}
\center{\includegraphics[width=1.1\linewidth]{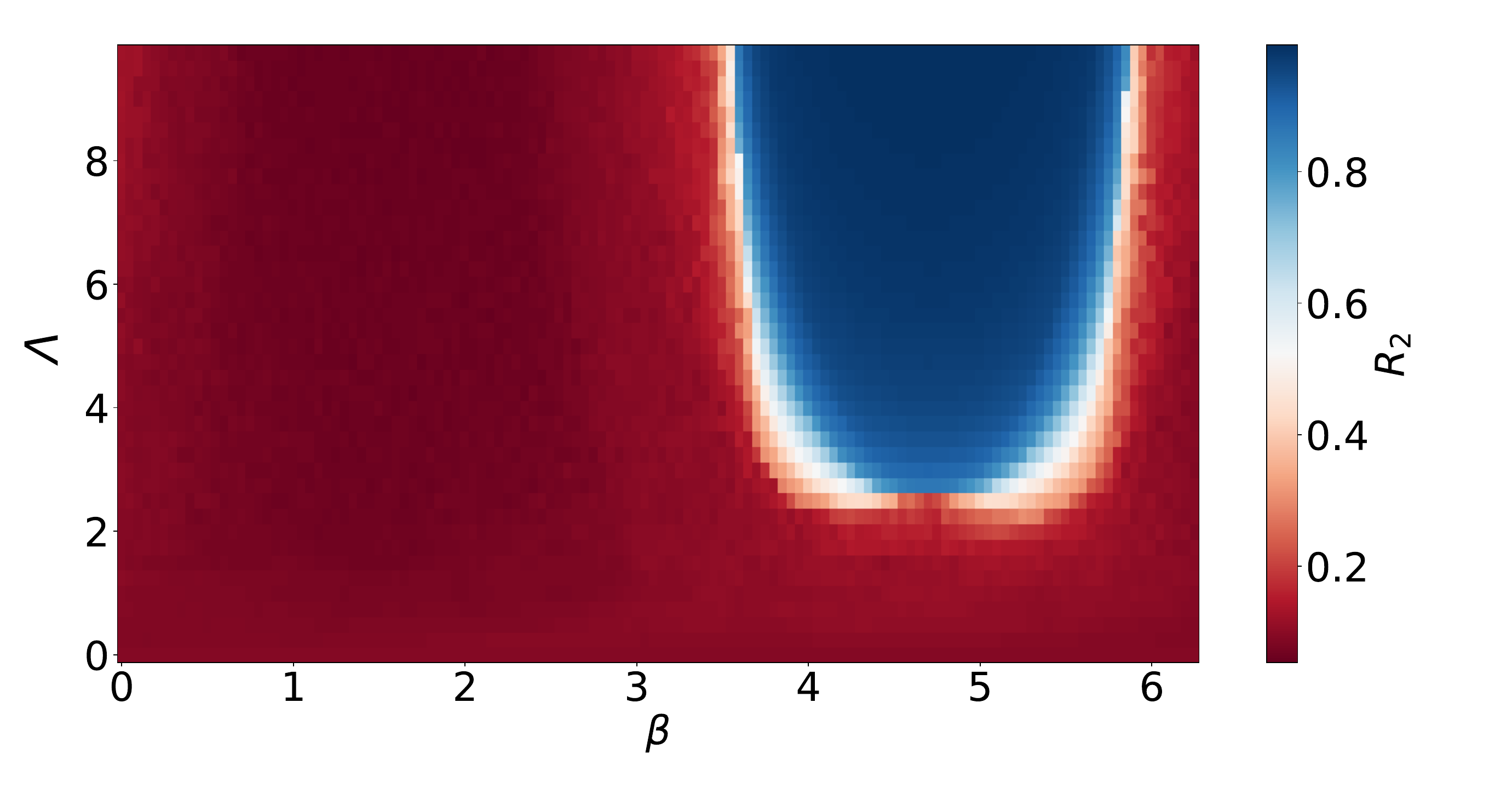}} \\ (b)
\end{minipage}
\caption{\label{fig:beta_K_R2} Order parameter $R_2$ depending on the parameters $(\beta, \Lambda)$. (a) Triadic couplings. (b) Pairwise couplings. }
\end{figure*}

In the mean-field approach \cite{ott2008,ott2009}, there is a limit transition to an infinite number of oscillators. In the discrete case, in the expression for the order parameter we have an averaging over the number of oscillators $R_m = \frac 1 N | \sum_{j=1}^N e^{i m\theta_j}|$. In the continuous case, averaging occurs over phases and frequencies:
\begin{equation}
R_m = \text{Re} \int_{locked} d\omega g(\omega) \int_0^{2\pi} d\theta f(\theta) e^{i m \theta}. \label{Rm_eq1}
\end{equation}

Since the distribution of the initial phases has a simple form
\begin{equation}
f(\theta) = \eta \delta(\theta) + (1-\eta)\delta(\theta-\pi), \label{theta_distribution}
\end{equation}
then for the stationary state this distribution has the form
\begin{equation}
f(\theta) = \eta \delta(\theta-\theta^*) + (1-\eta) \delta(\theta-\theta^*-\pi). \label{theta_distribution_stat}
\end{equation}

Therefore, 
\begin{eqnarray}
&& R_1 = (2\eta-1) \int_{locked} d\omega g(\omega) \cos(\theta^*), \notag \\
&& R_2 = \int_{locked} d\omega g(\omega) \cos(2 \theta^*), \label{Rm_eq2}
\end{eqnarray}
where $g(\omega) = \frac {1}{2\Delta}$. Taking the integrals for a stable stationary state, we obtain
\begin{eqnarray}
&& R_1 = \frac {(2\eta-1) \Lambda R_2^2}{2\Delta}  \notag \\ && \times \Bigl \{  \frac 1 3 \sin \Bigl [ \frac 3 4 \text{arccos} \Bigl (  \frac{-2\Delta+\Lambda\cos\beta}{\Lambda R_2^2} \Bigr ) - \frac {\beta}{4} \Bigr ] \notag \\ && -\frac 1 5 \sin \Bigl[ \frac 5 4 \text{arccos} \Bigl ( \frac{-2\Delta+\Lambda\cos\beta}{\Lambda R_2^2} \Bigr ) + \frac {\beta}{4} \Bigr] \notag \\ && -\frac 1 3 \sin \Bigl [ \frac 3 4 \text{arccos} \Bigl ( \frac{2\Delta+\Lambda\cos\beta}{\Lambda R_2^2} \Bigr ) - \frac {\beta}{4} \Bigr] \notag \\ && +\frac 1 5 \sin \Bigl[ \frac 5 4 \text{arccos} \Bigl ( \frac{2\Delta+\Lambda\cos\beta}{\Lambda R_2^2} \Bigr ) + \frac {\beta}{4} \Bigr]  \Bigr \}, \label{eq_R1_theta1_3D} \\
&& R_2 = \frac {\Lambda R_2^2} {4\Delta} \Bigl\{ \cos \Bigl[ -\frac 1 2 \text{arccos} \Bigl ( \frac{-2\Delta+\Lambda\cos\beta}{\Lambda R_2^2} \Bigr ) + \frac {\beta}{2} \Bigr] \notag \\ && +\frac 1 3 \cos\Bigl[ \frac 3 2 \text{arccos} \Bigl( \frac{-2\Delta+\Lambda\cos\beta}{\Lambda R_2^2} \Bigr) + \frac {\beta}{2} \Bigr] \notag \\ && -\cos \Bigl[ -\frac 1 2 \text{arccos} \Bigl ( \frac{2\Delta+\Lambda\cos\beta}{\Lambda R_2^2} \Bigr ) + \frac {\beta}{2} \Bigr] \notag \\ && -\frac 1 3 \cos \Bigl[ \frac 3 2 \text{arccos} \Bigl( \frac{2\Delta+\Lambda\cos\beta}{\Lambda R_2^2} \Bigr) + \frac {\beta}{2} \Bigr] \Bigr\} \label{eq_R2_theta1_3D}.
\end{eqnarray}

It is worth noting that the procedure for obtaining expressions for $R_1$, $R_2$ in the case of pairwise couplings is carried out in a similar way.

\section{\label{sec:numerical} Numerical results}

\begin{figure*}
\begin{minipage}[h]{0.47\linewidth}
\center{\includegraphics[width=1.1\linewidth]{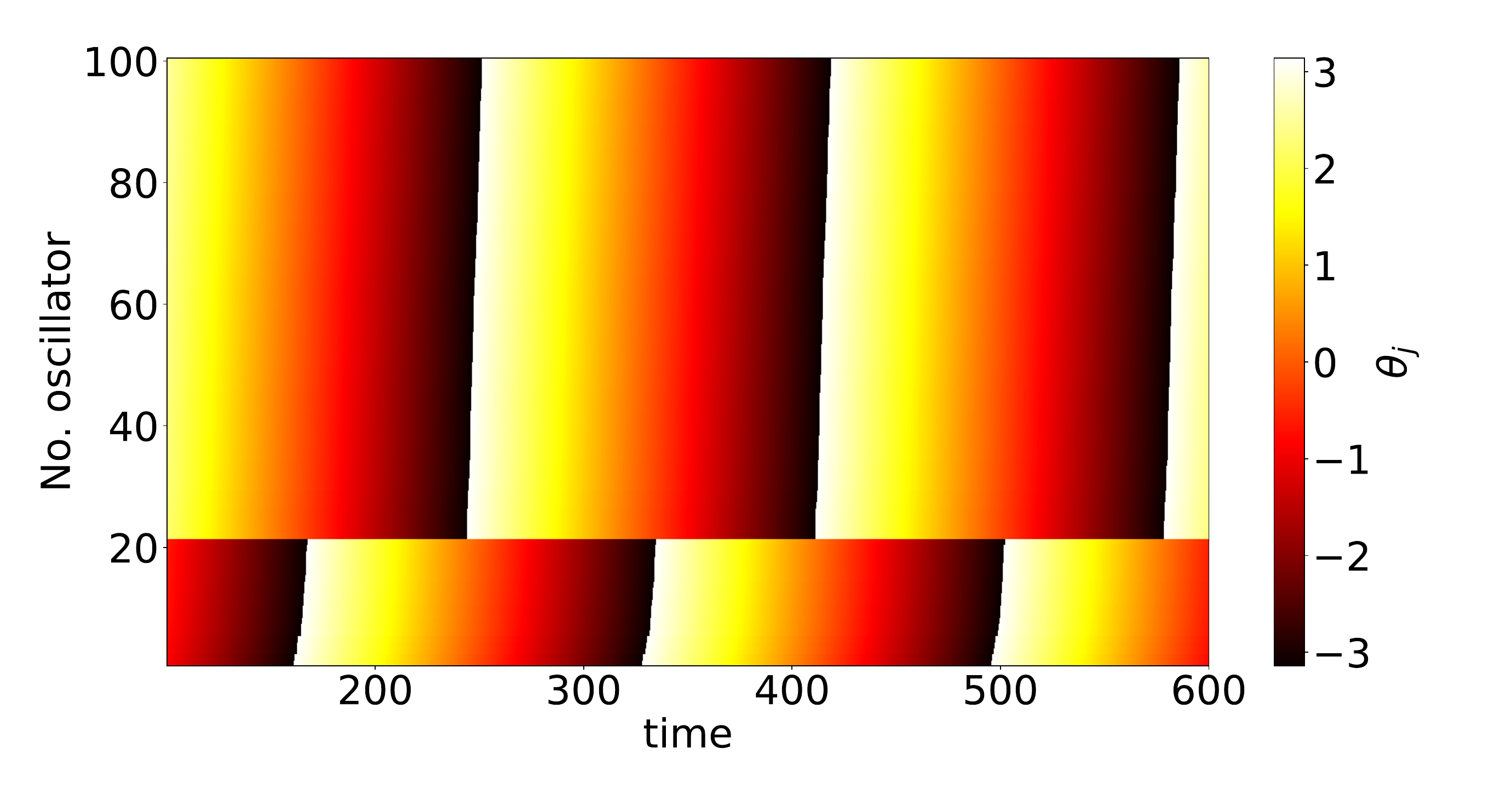}} \\
\end{minipage}
\hfill
\begin{minipage}[h]{0.47\linewidth}
\center{\includegraphics[width=1.1\linewidth]{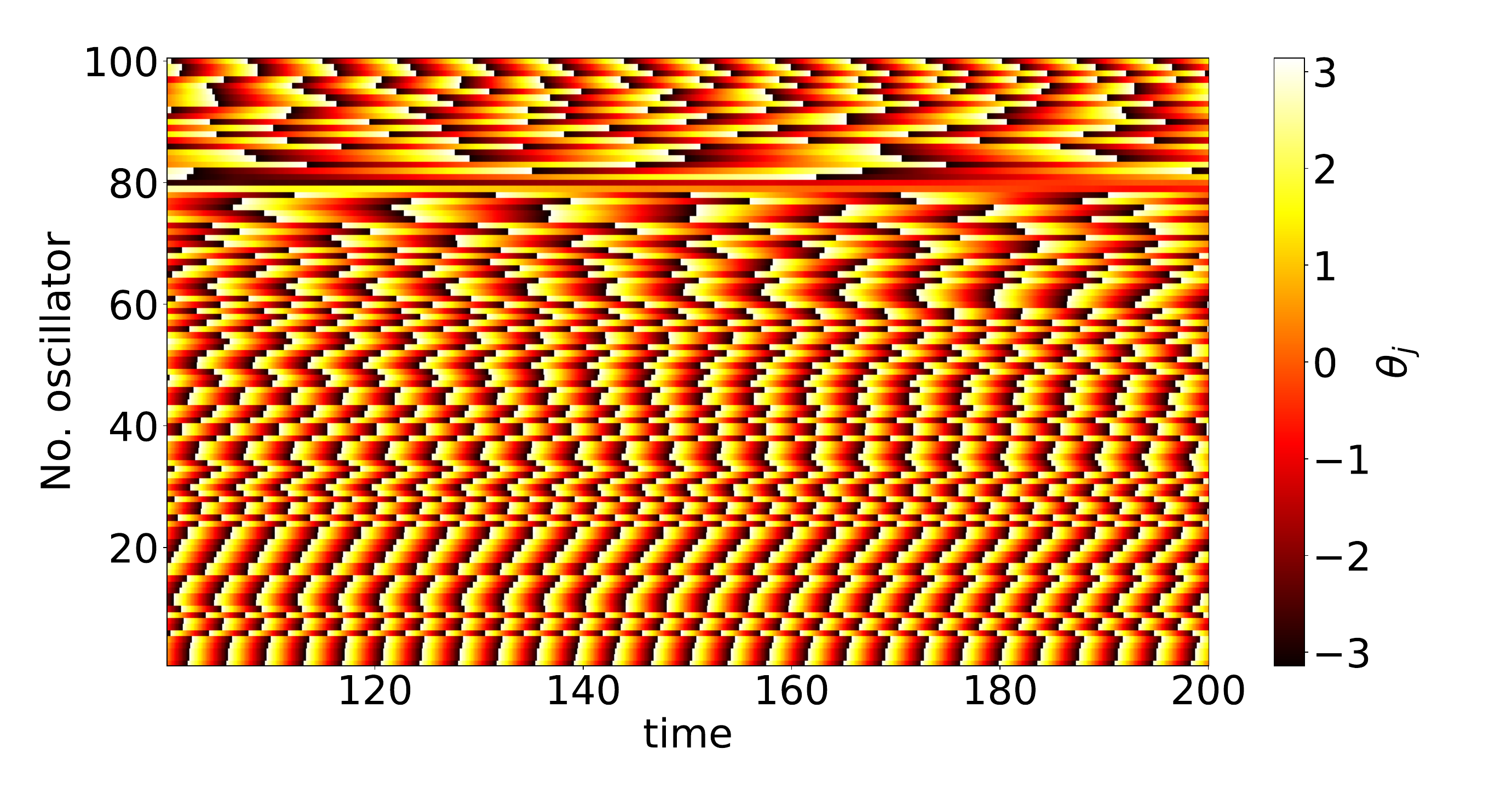}} \\
\end{minipage}
\vfill
\begin{minipage}[h]{0.47\linewidth}
\center{\includegraphics[width=1.1\linewidth]{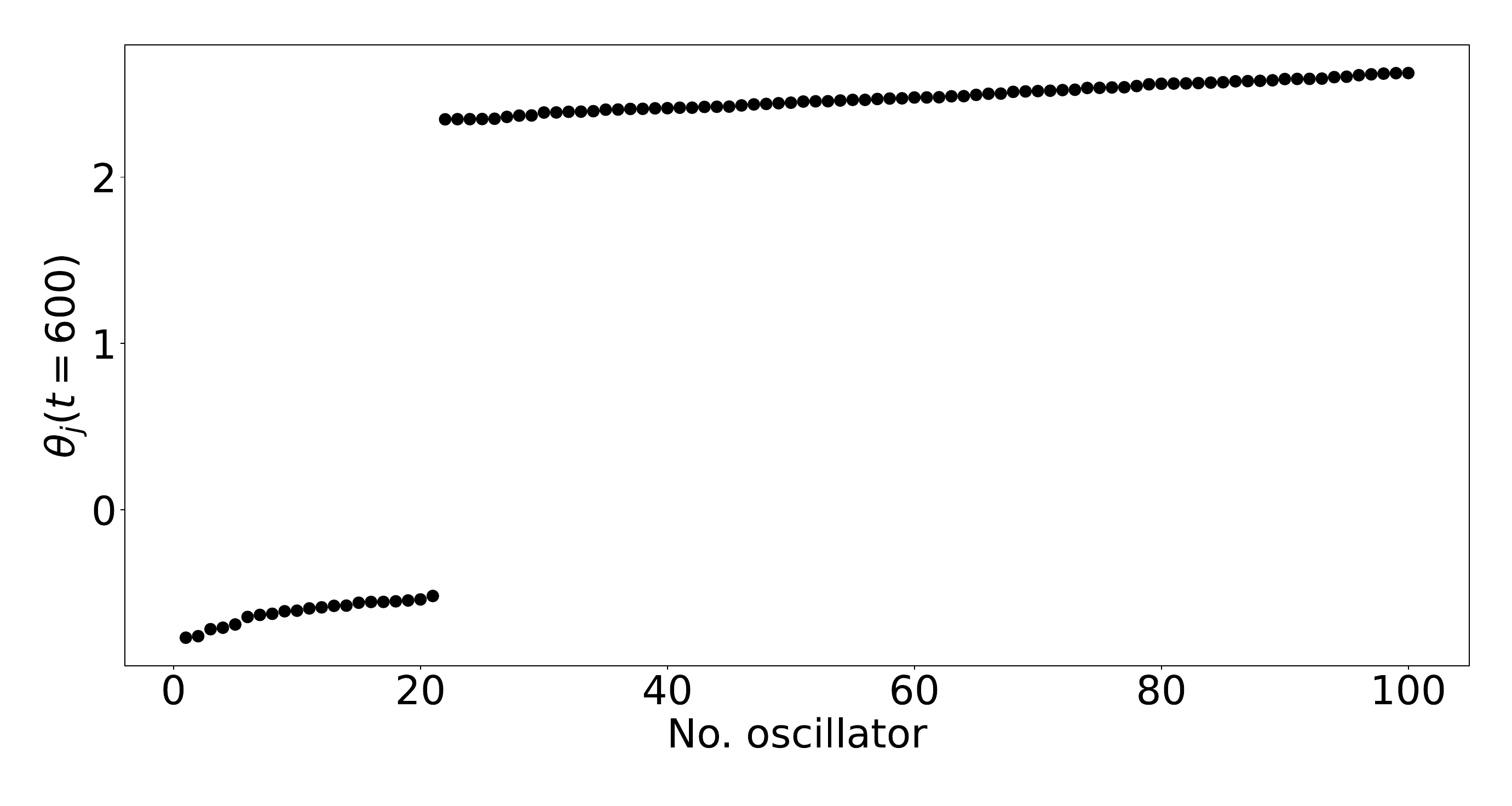}} \\
\end{minipage}
\hfill
\begin{minipage}[h]{0.47\linewidth}
\center{\includegraphics[width=1.1\linewidth]{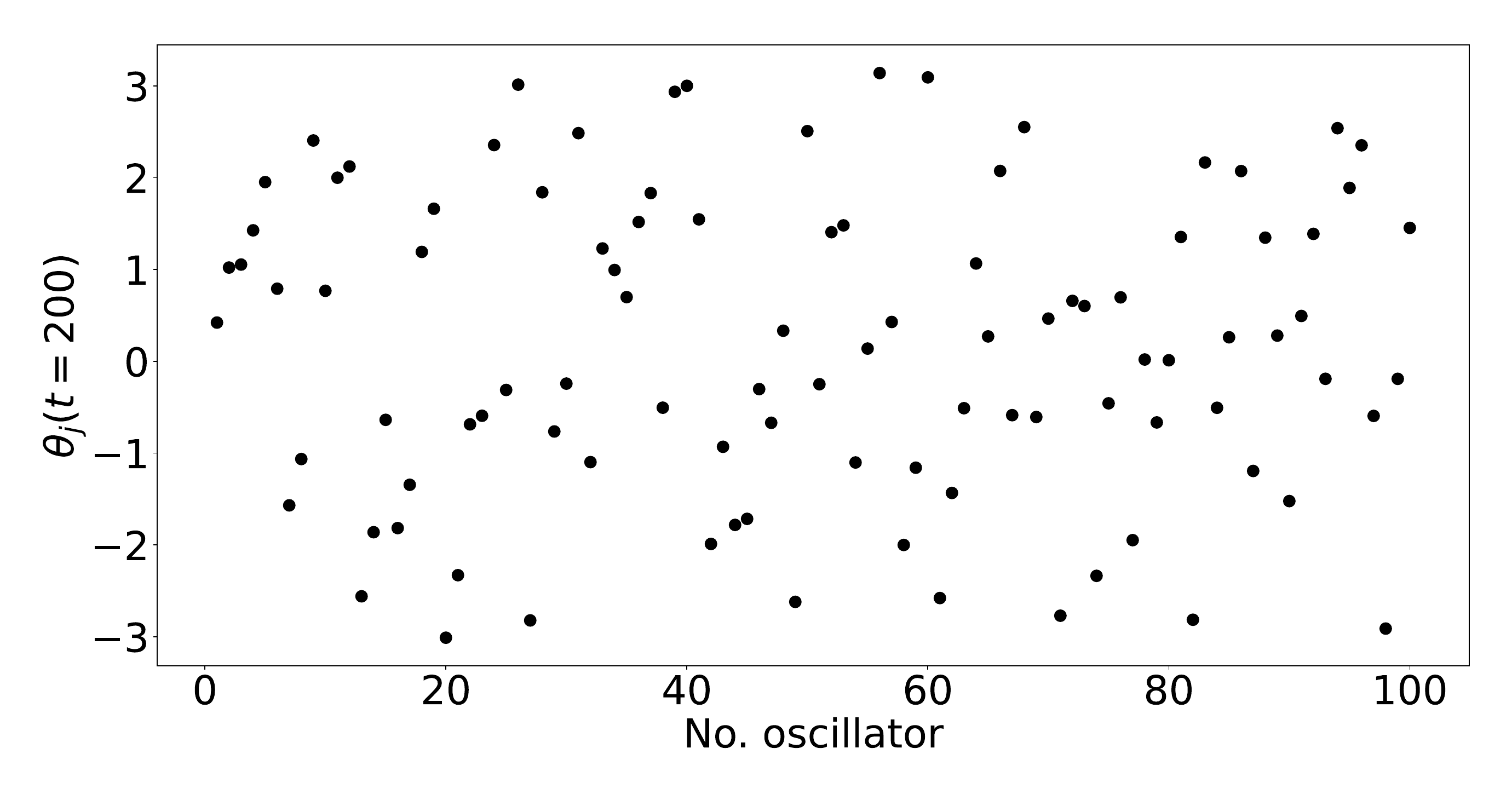}} \\
\end{minipage}
\vfill
\begin{minipage}[h]{0.47\linewidth}
\center{\includegraphics[width=1.1\linewidth]{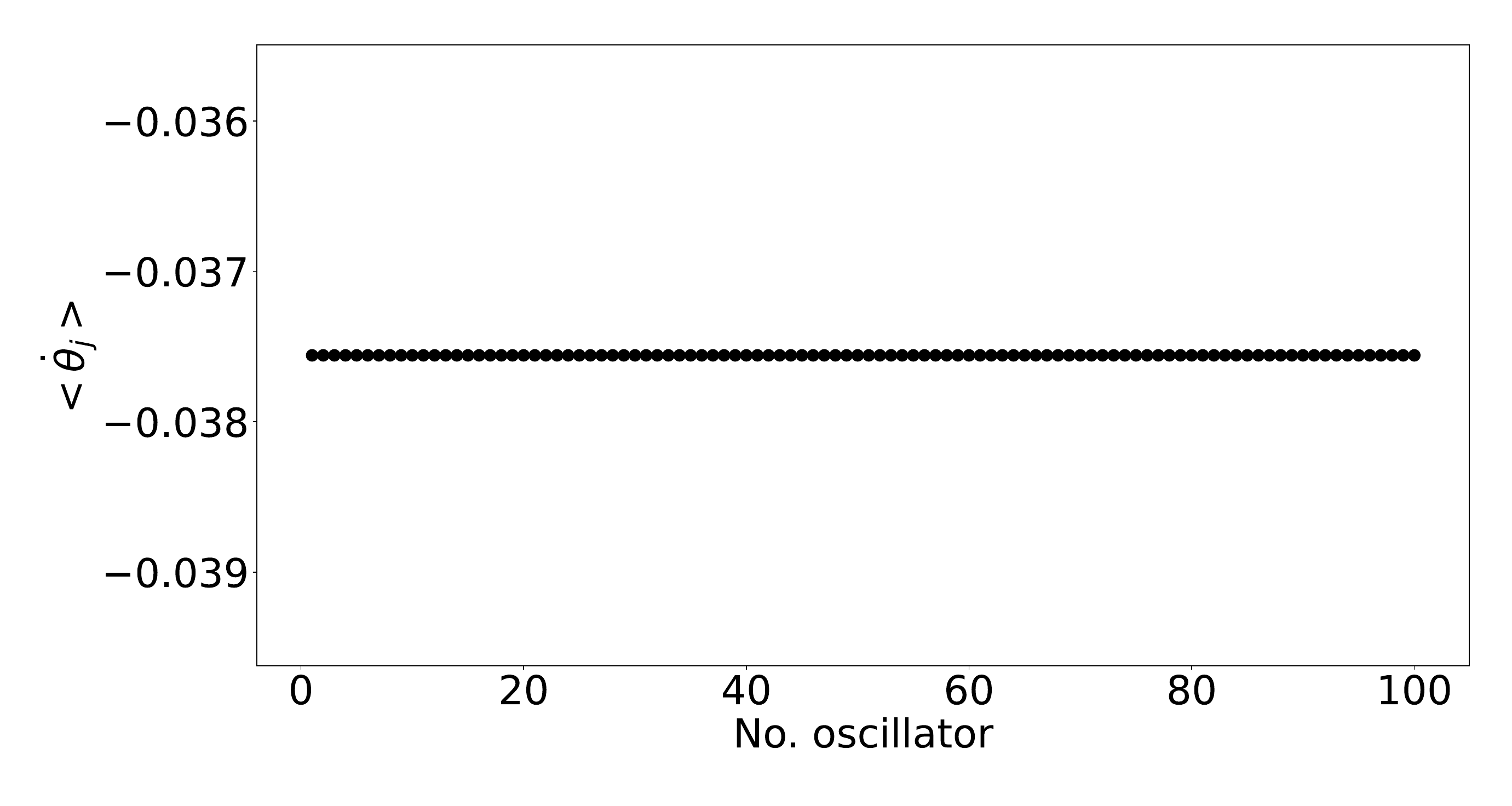}} \\
\end{minipage}
\hfill
\begin{minipage}[h]{0.47\linewidth}
\center{\includegraphics[width=1.1\linewidth]{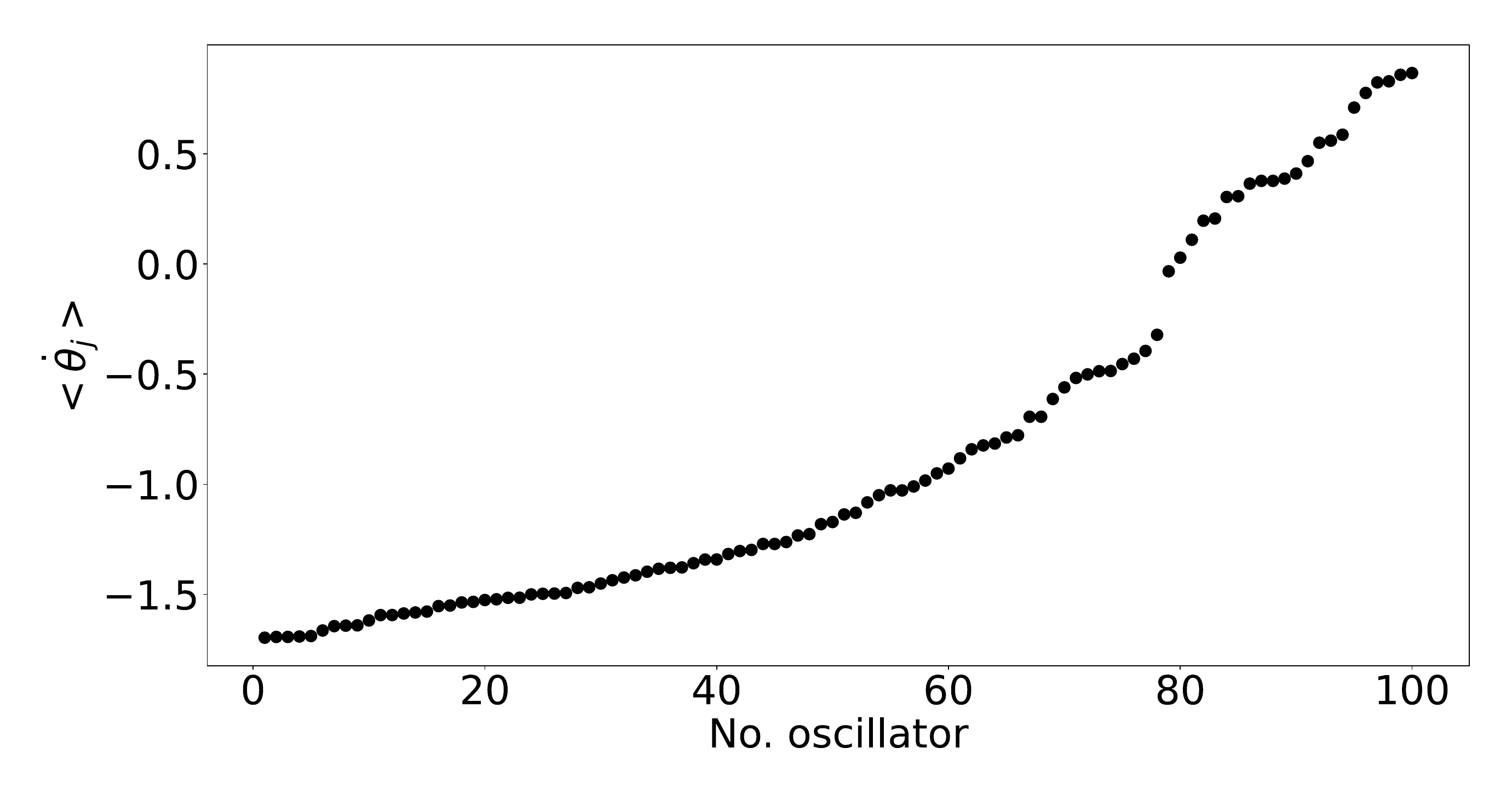}} \\
\end{minipage}
\vfill
\begin{minipage}[h]{0.47\linewidth}
\center{\includegraphics[width=1.1\linewidth]{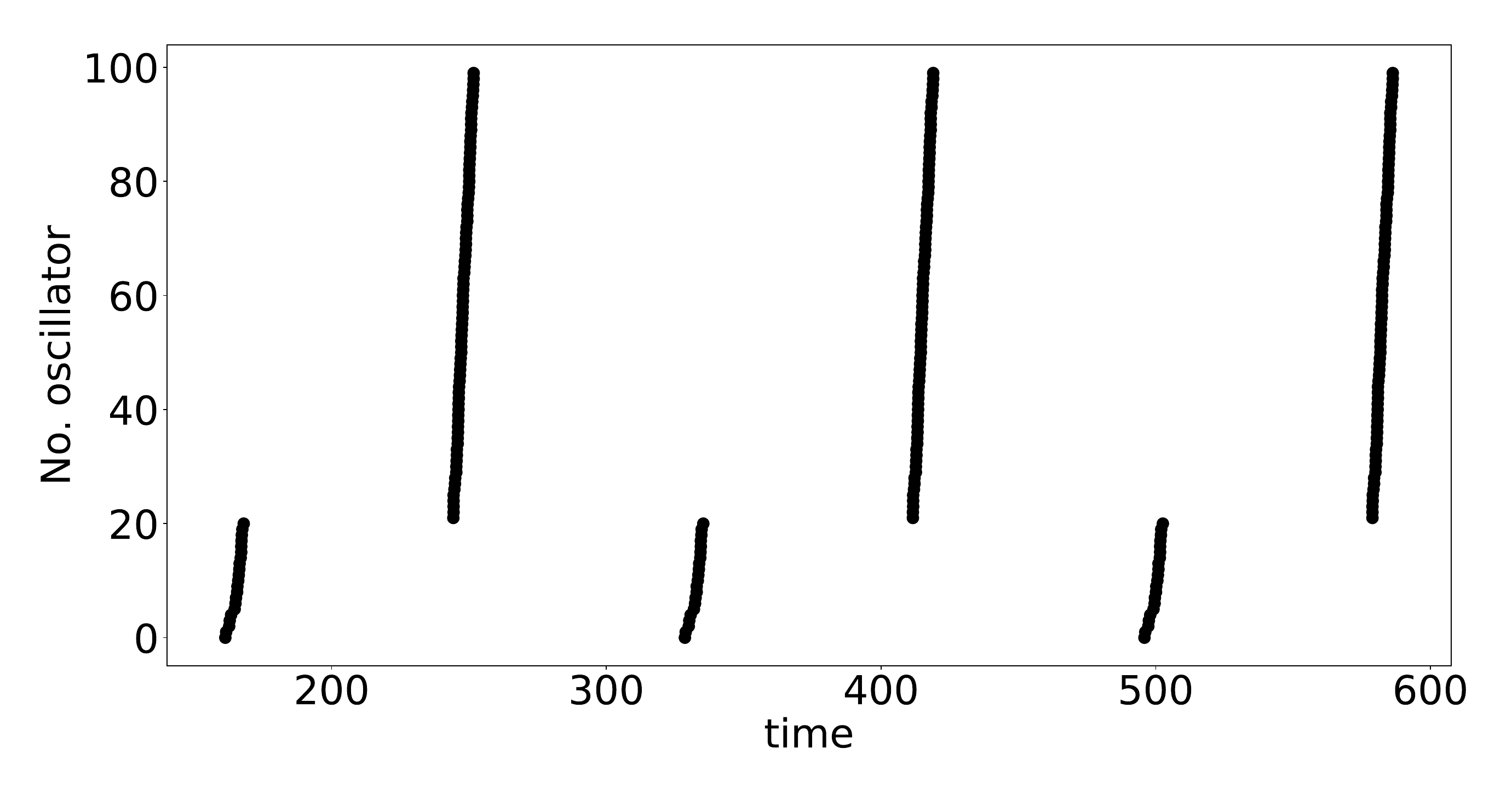}} \\ (a)
\end{minipage}
\hfill
\begin{minipage}[h]{0.47\linewidth}
\center{\includegraphics[width=1.1\linewidth]{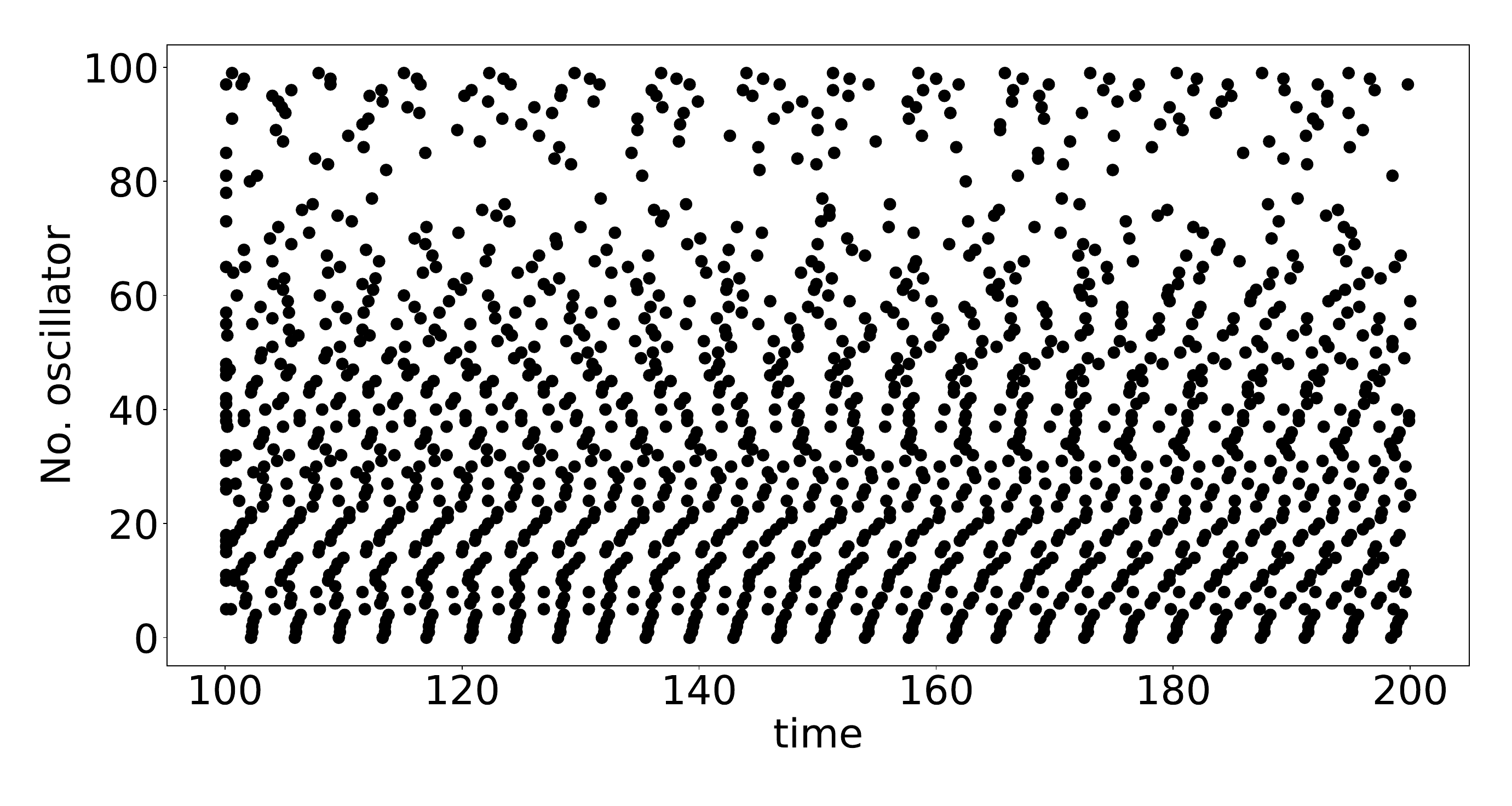}} \\ (b)
\end{minipage}
\caption{\label{fig:space-time} The first line shows the spatial-temporal diagrams. The second line shows the distributions of the phases at some moment in time. The third line shows the distributions of the time-averaged frequencies of the oscillators. The forth line shows the spike rasters. \\ (a) Synchronous state. $\beta = 5, \; \eta=0.8$. (b)  Asynchronous state. $\beta = 0, \; \eta=0.8$. }
\end{figure*}

\begin{figure}
\includegraphics[width=1\linewidth] {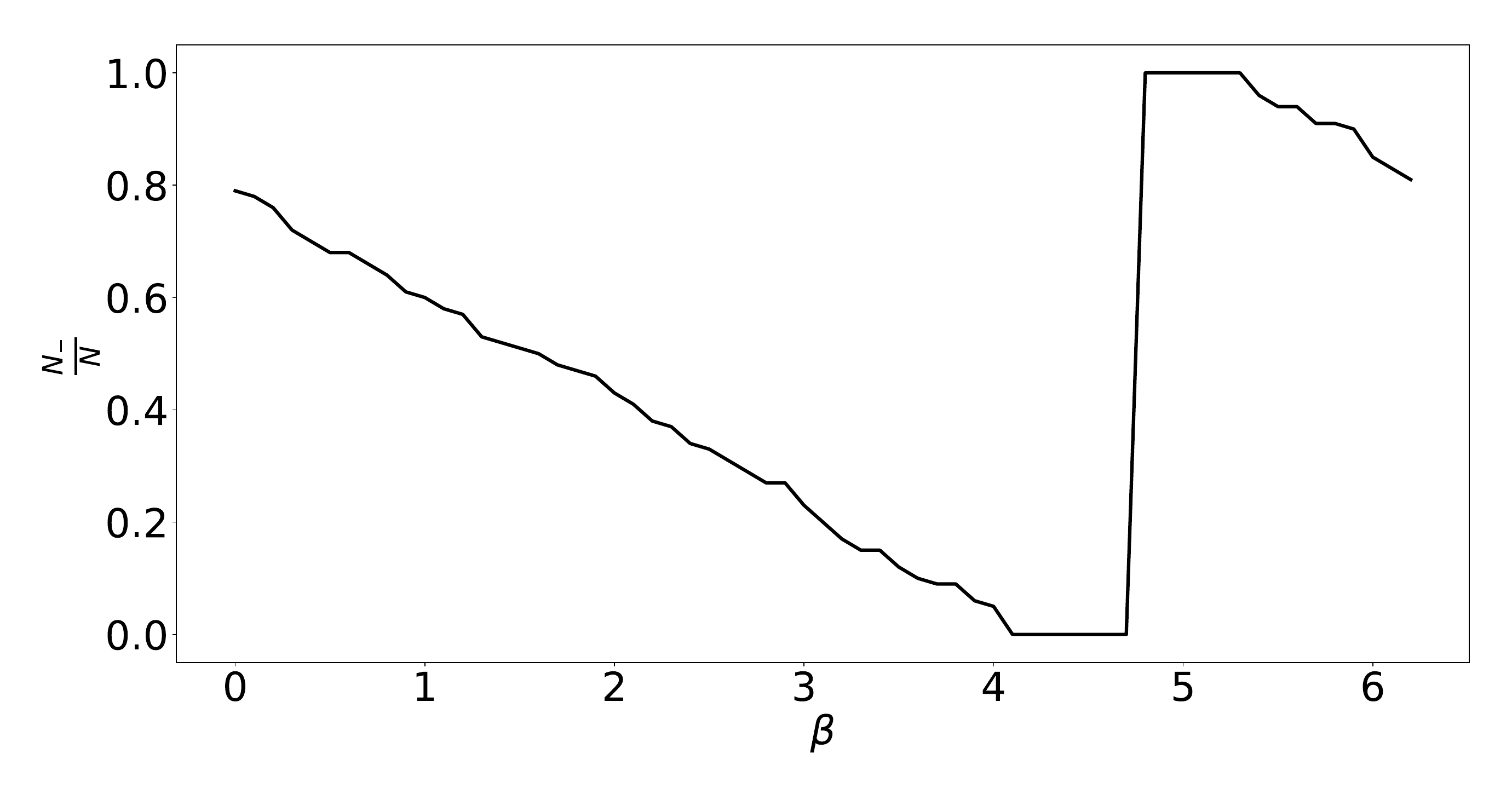}
\caption{\label{fig:beta_percent} Fraction of the oscillators with negative frequencies depending on  the parameter $\beta$ for $\eta = 0.8$. }
\end{figure}

We carried out numerical simulation of the network dynamics using the four-order Runge-Kutta method. Then we have studied the quantities that characterize the degree of synchronization, depending on the parameters of the system. Such quantities are the order parameters $R_1$ and $R_2$, as well as the proportion of synchronized pairs of oscillators between which there is a connection $R_{link}$ and the proportion of network elements that are synchronized with at least one other element $P_c$. The last two quantities can be entered as follows \cite{Kasatkin_2018, Kasatkin_2023}. First, we introduce the quantity $R_{ij} = |\lim_{t \rightarrow \infty} \frac {1}{\Delta t} \int_{T}^{T+\Delta t} e^{i(\theta_i(t) - \theta_j(t))} dt|$, which characterizes the degree of synchronization between arbitrary oscillators $i$ and $j$ of the network. Then we assign $\tilde {R_{ij}} = 1$ if $R_{ij} = 1$ and $\tilde {R_{ij}} = 0$ if $R_{ij} < 1$. And then we introduce the quantities $R_{link}$ and $P_c$ according to the formulas $R_{link} = \frac {\sum_{i,j=1}^N A_{ij} \tilde {R_{ij}}} {2 \sum_{i,j=1}^N A_{ij}}$ and $P_c = \frac {1}{N} \sum_{j=1}^N \max_{i \neq j} \tilde{R_{ij}}$, where $A_{ij}$ are the elements of the two-dimensional adjacency matrix.

Figure \ref{fig:beta_Rm} shows the order parameters $R_1$, $R_2$ and $R_{link}$, $P_c$ as a function of the parameter $\beta$ for $\eta=0.8$ in system \eqref{sys_3D} (FIG. \ref{fig:beta_Rm}a) and in system \eqref{sys_2D} (FIG. \ref{fig:beta_Rm}b). The dots show the results of the calculation, the lines show the theoretical curves from the expressions for the order parameters. The theory fits the numerical experiment. Figure \ref{fig:beta_Rm}a demonstrates, firstly, that in system \eqref{sys_3D}, there is a certain interval within which $R_2$ is close to $1$, $R_{link}$ and $P_c$ are equal to $1$, and $R_1$ takes a value between $0$ and $1$. This means that in this interval of the parameter $\beta$, there are two synchronous clusters in the system, shifted by $\pi$ relative to each other. This interval is located around $\beta = \frac {3\pi}{2}$, which corresponds to the Hebbian plasticity rule. Secondly, only two states are realized in system \eqref{sys_3D}: synchronous and asynchronous. When the parameter $\beta$ varies, an abrupt jump in the order parameters occurs. This is the fundamental difference between triadic and pairwise couplings. For comparison, FIG. \ref{fig:beta_Rm}b shows the case of pairwise couplings \eqref{sys_2D}, when a smooth change in the order parameters occurs. It is accompanied by the appearance of chimera states, when some oscillators remain in the synchronous group, while others separate from it \cite{kas2017, Kasatkin_2018, Kasatkin_2023}. Thus, we find out an abrupt transition of the order parameters with a variation of the parameter $\beta$ in \eqref{sys_3D}. Also, it is worth noting that the parameter $\eta$ enters the expression for $R_1$ \eqref{eq_R1_theta1_3D} linearly, affecting the height of the plateau in the synchronous state. At $\eta = 1$ its height is maximal and close to $1$, at $\eta=0.5$ its height is minimal and equal to zero. At the same time, the parameter $\eta$ has no effect on the height of the plateau of $R_2$: since we have just two different initial conditions for phases, there are always two clusters in the system \eqref{sys_3D} in the synchronous state (except for the case $\eta=1$, when there is one cluster), and always $R_2 \approx 1$.

Figure \ref{fig:beta_K_R2}a demonstrates the dependence of the order parameter $R_2$ on the parameters $(\beta, \Lambda)$ in system \eqref{sys_3D}. As one can see, there is a region (shown in blue) where $R_2 \approx 1$, i.e., where there are two synchronous clusters in the system \eqref{sys_3D}. There is a synchronization threshold $\Lambda \approx 2.1$, from which this region begins. As the parameter $\Lambda$ increases, this area gradually expands in terms of the parameter $\beta$. It is worth noting that the dependences of $R_1$, $R_{link}$ and $P_c$ on the parameters $(\beta, \Lambda)$ look approximately the same -- they have exactly the same area where these values are close to $1$ or are equal to $1$. Also from FIG. \ref{fig:beta_K_R2}a, we again see that an abrupt jump in the order parameters occurs when both the parameter $\Lambda$ and the parameter $\beta$ are varied when moving from the region of synchronization to the region of desynchronization. For comparison, FIG. \ref{fig:beta_K_R2}b demonstrates that in the case of pairwise couplings there is a smooth transition. 

Figure \ref{fig:space-time} shows the dynamics of the system \eqref{sys_3D} in synchronous and asynchronous states. When constructing FIG. \ref{fig:space-time}, we ordered the oscillators as follows: first, by increasing time-averaged frequencies, and then by increasing phases within each cluster. The first line of FIG. \ref{fig:space-time} shows the spatial-temporal diagrams, where color means the phase of each oscillator. As one can see, with given parameter values, in the synchronous state there are two clusters in the system \eqref{sys_3D} shifted relative to each other by $\pi$. The size of each cluster depends on the parameter $\eta$: at $\eta = 0.5$ the clusters are the same in size, at $\eta \approx 0.99$ the size of one cluster will be maximum, and the second -- minimum. In the asynchronous state, there is a spatial-temporal disorder. The second line of FIG. \ref{fig:space-time} demonstrates the distributions of the phases at some moment in time. As one can see, in the asynchronous state the phases are randomly distributed, while in the synchronous state they are clustered around some two values. The third line shows the distributions of the time-averaged frequencies of the oscillators. In the synchronous state, both clusters have the same frequency, while in the asynchronous state, the frequencies are distributed in a certain interval. The forth line shows the spike rasters. That is, if we consider each oscillator as a neuron, then each point in spike raster corresponds to the spike of a particular neuron, i.e., the moment when its phase crosses the value of $\pi$. In the synchronous state, the oscillators spike almost simultaneously within each cluster, while in the asynchronous state, they spike randomly.

\begin{figure}
\includegraphics[width=1\linewidth] {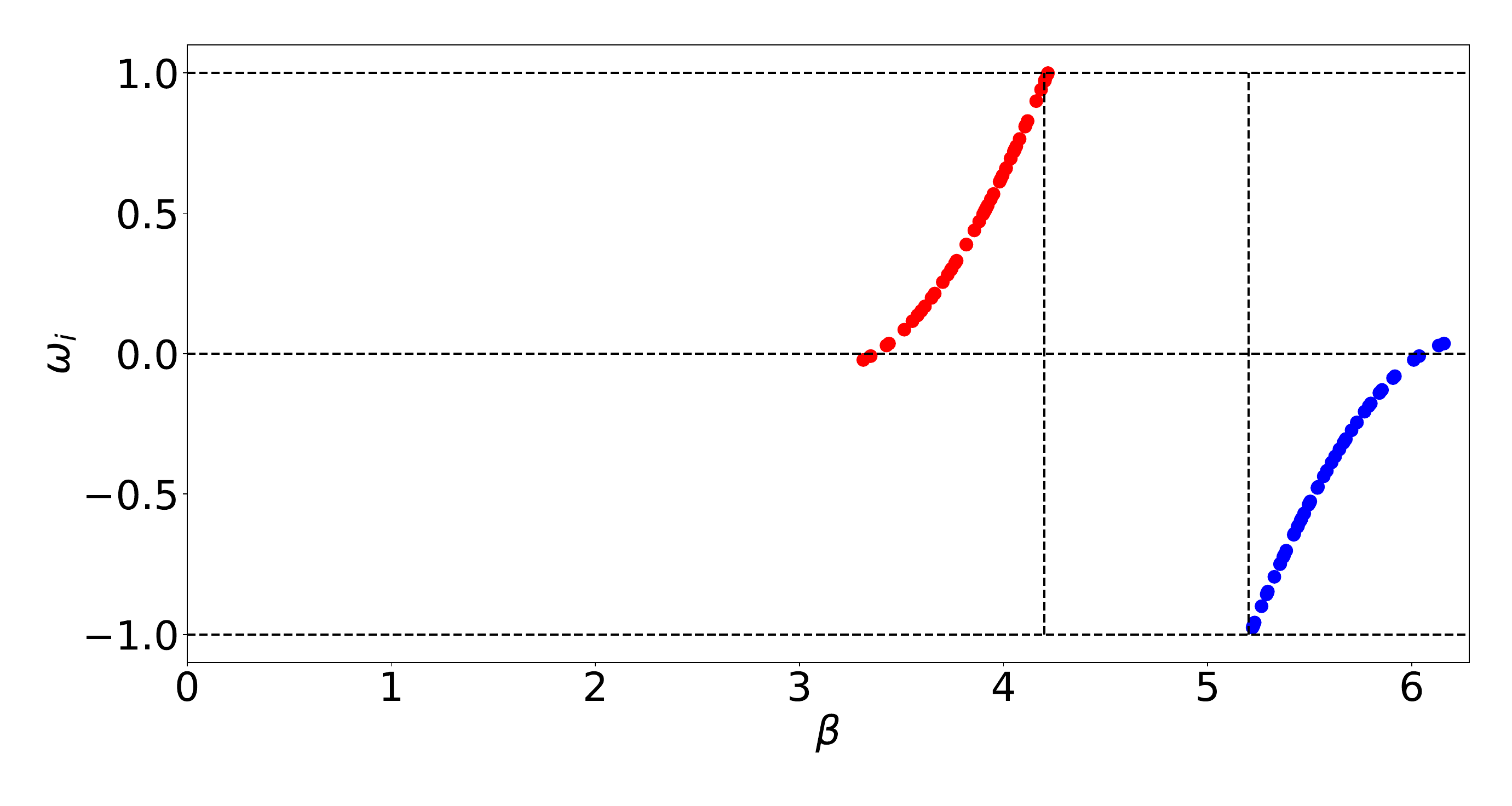}
\caption{\label{fig:beta_omegas_threshold} Dependence of the threshold values of the parameter $\beta$ on the natural frequencies $\omega_i$ for $R_2 = 0.987$ corresponding to the synchronous state. Red dots correspond to the left border and blue dots correspond to the right border. The network synchronization interval is located between the two vertical dotted lines.}
\end{figure}

\begin{figure*}
\begin{minipage}[h]{0.47\linewidth}
\center{\includegraphics[width=1.1\linewidth]{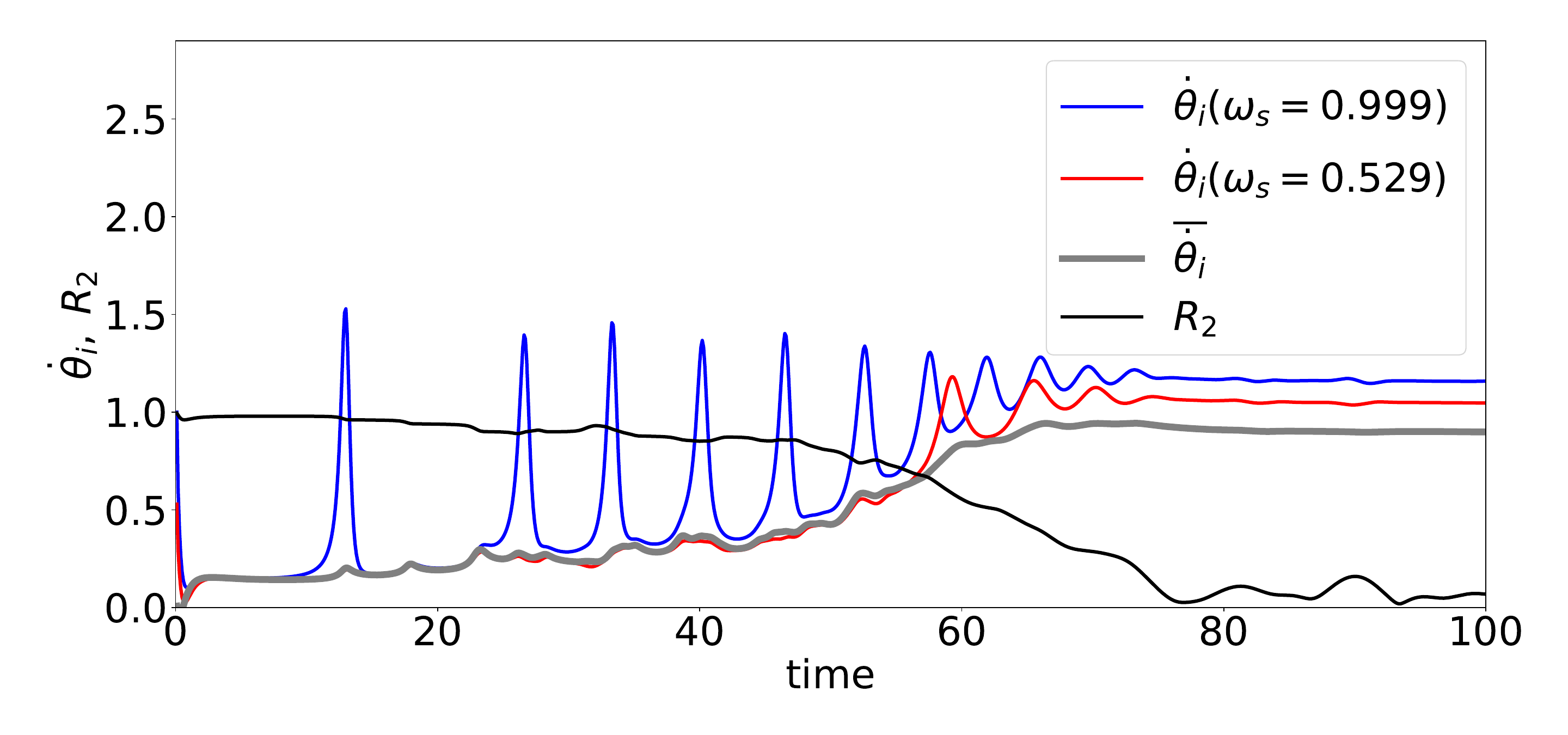}} \\ (a)
\end{minipage}
\hfill
\begin{minipage}[h]{0.47\linewidth}
\center{\includegraphics[width=1.1\linewidth]{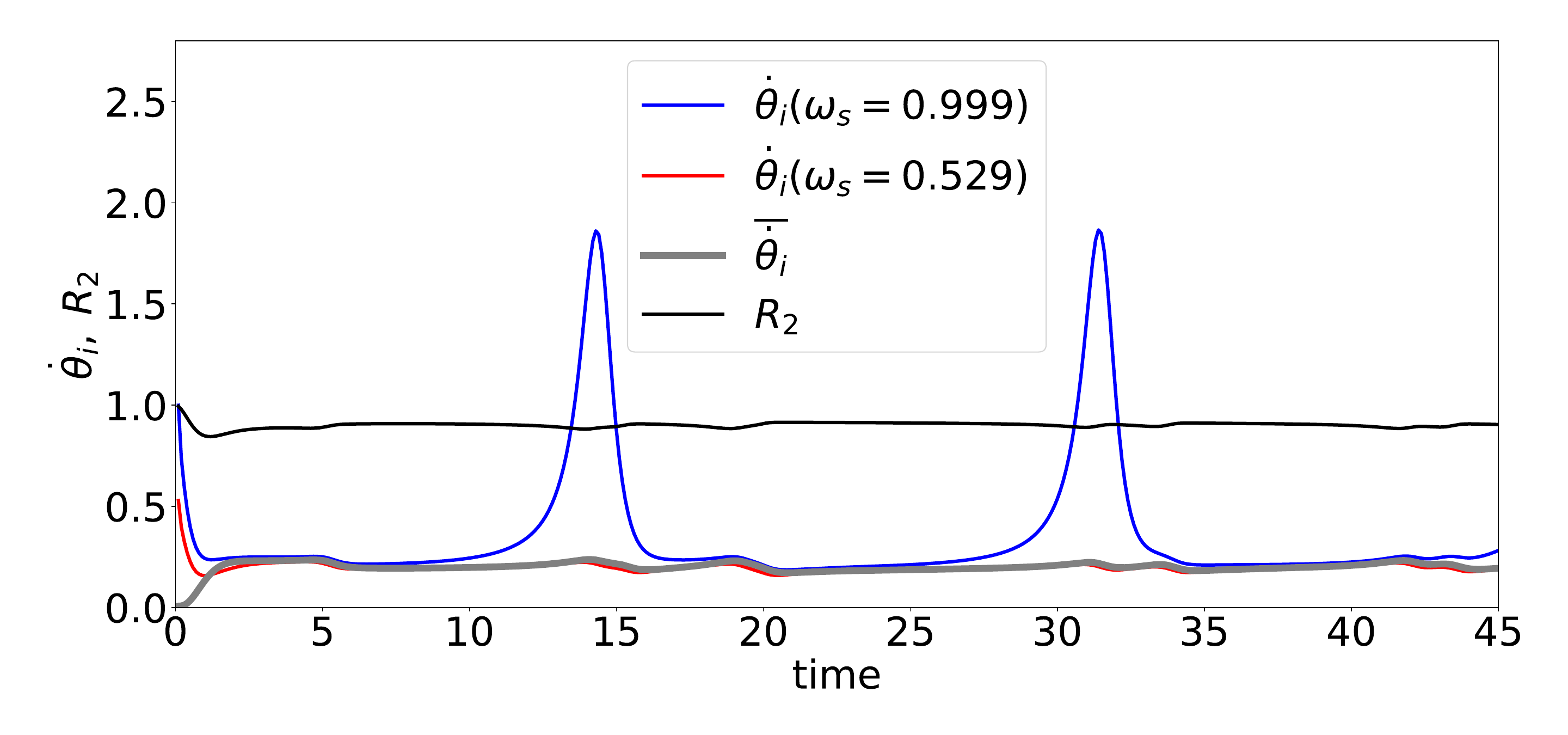}} \\ (b)
\end{minipage}
\caption{\label{fig:time_freqs} Frequencies of the oscillator with the highest natural frequency (blue), the oscillator with the 25th natural frequency (red), the average network frequency (gray) and the order parameter $R_2$ (black) vs. time during desynchronization. $\beta=4.15, \; \eta=0.8$. \\ (a) Triadic couplings. (b) Pairwise couplings.}
\end{figure*}

\begin{figure*}
\begin{minipage}[h]{0.47\linewidth}
\center{\includegraphics[width=1.1\linewidth]{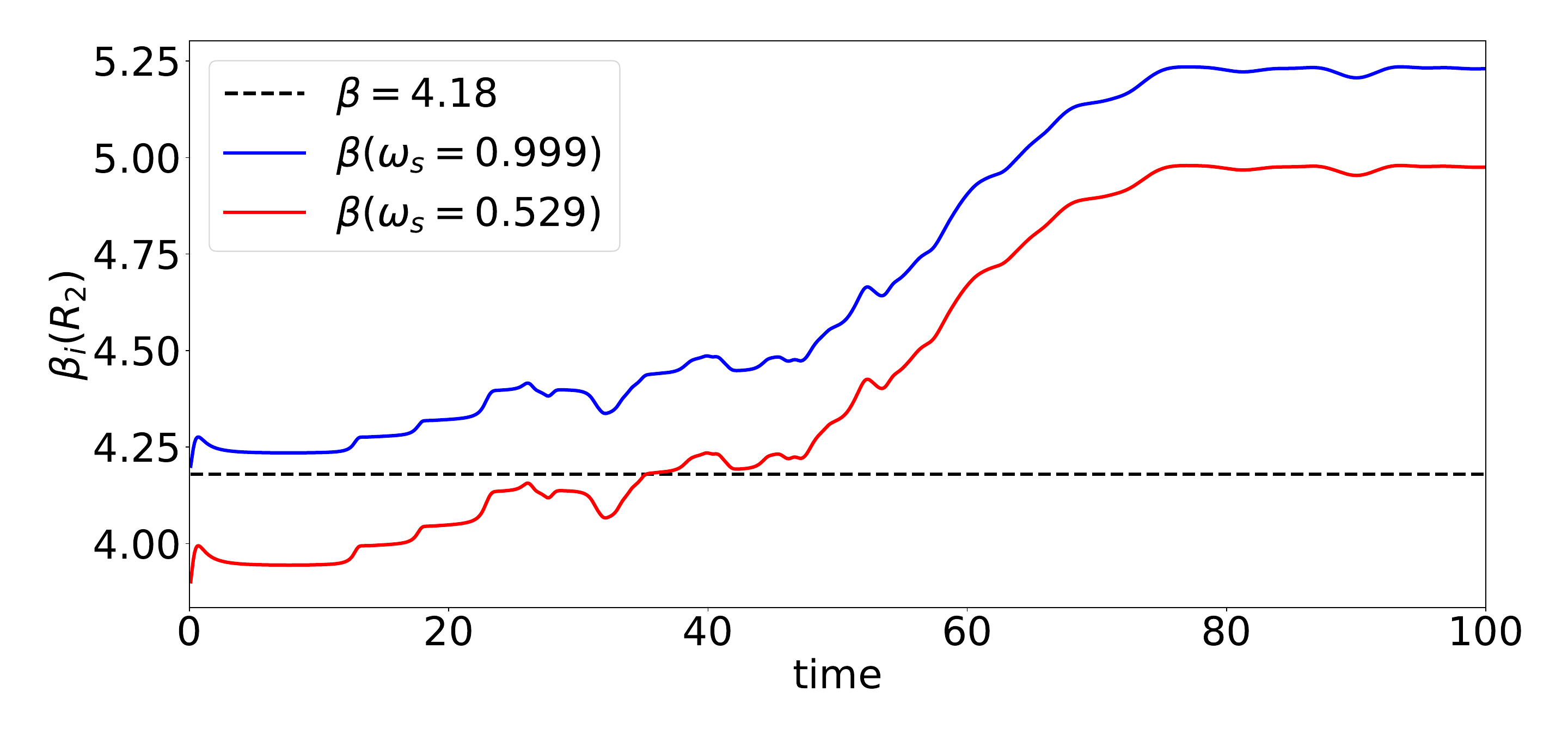}} \\ (a)
\end{minipage}
\hfill
\begin{minipage}[h]{0.47\linewidth}
\center{\includegraphics[width=1.1\linewidth]{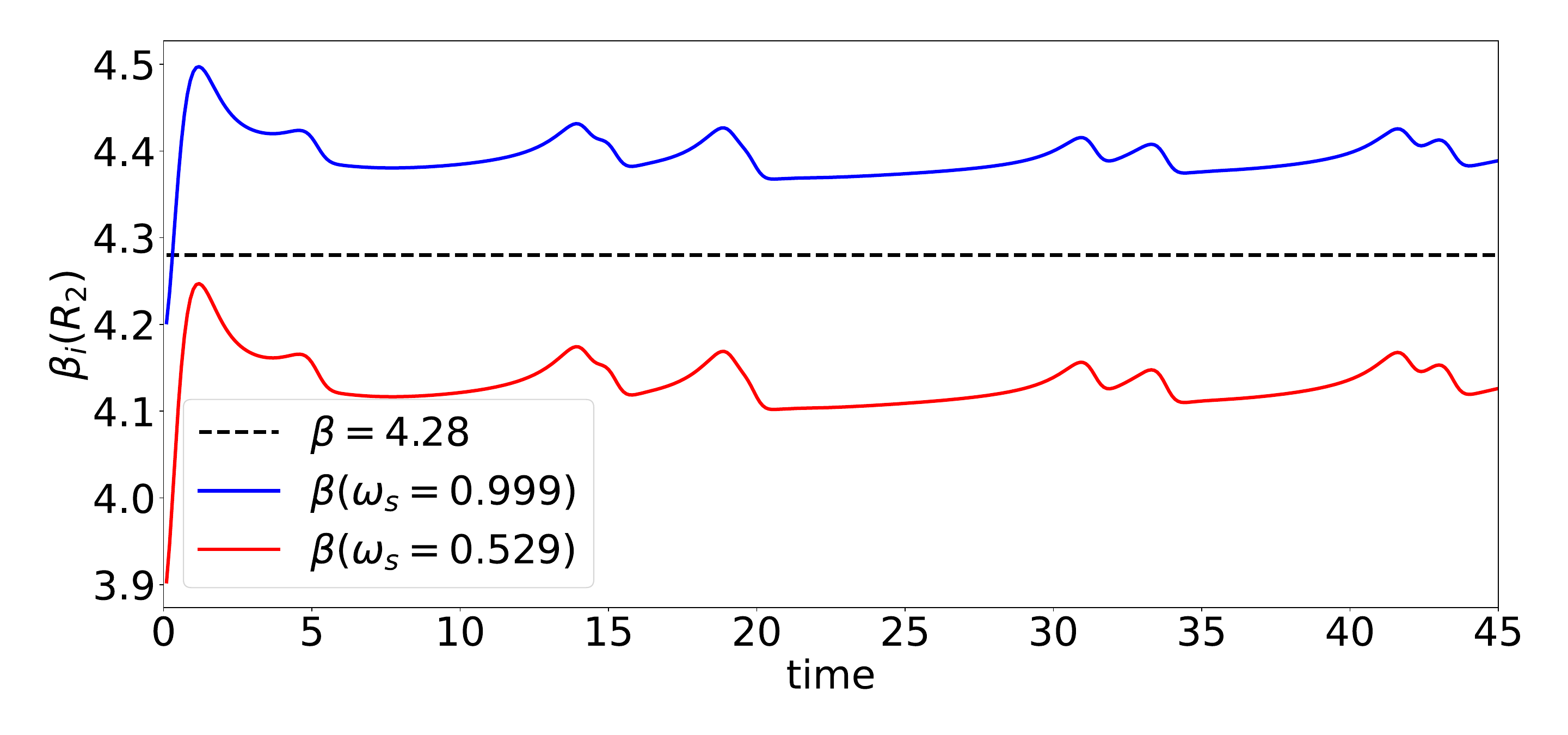}} \\ (b)
\end{minipage}
\caption{\label{fig:time_threshold} The dependence of the threshold values $\beta_{left}$ for the oscillator with the highest natural frequency (blue) and for the oscillator with the 25th natural frequency (red). The dotted horizontal line shows the boundary of the network synchronization interval. $\beta=4.15, \; \eta=0.8$. \\ (a) Triadic couplings. (b) Pairwise couplings. }
\end{figure*}

As one can see from FIG. \ref{fig:space-time}, the time-averaged oscillator frequencies take both positive and negative values. Parameters affect the frequency value. We have plotted the dependence of the fraction of negative frequencies on the parameter $\beta$ (FIG. \ref{fig:beta_percent}). This fraction changes, and the synchronous interval is divided in two -- in one half the common frequency is positive, and in the other it is negative.

\section{\label{sec:loss} Loss of stability}

Consider the process of loss of stability of the synchronous state of the network. From the condition of the double root $\theta_{i}^{1*} = \theta_{i}^{2*}$, we can obtain an expression for the value of the parameter $\beta$, at which the oscillator with index $i$ loses stability:
\begin{eqnarray}
&& \beta_{left} = 2\pi - \text{arccos} \Bigl [ \frac {2}{\Lambda} \Bigl (\omega_i - \frac{\Lambda R_2^2}{2} \Bigr )  \Bigr ], \label{left_border_3D} \\
&& \beta_{right} = 2\pi -\text{arccos} \Bigl [ \frac {2}{\Lambda} \Bigl (\omega_i + \frac{\Lambda R_2^2}{2} \Bigr )  \Bigr ]. \label{right_border_3D}
\end{eqnarray}

These thresholds values of $\beta$ for each network natural frequency $\omega_i$ are shown in FIG. \ref{fig:beta_omegas_threshold}, where the network synchronization interval is located between the two vertical dotted lines. Note that a part of the natural frequencies is not shown in FIG. \ref{fig:beta_omegas_threshold}, since these eigenfrequencies do not satisfy the domain of definition of the $\text{arccos}$ function. Closest to the boundaries of the synchronization region are the high-frequency oscillators: the oscillator with the maximal positive natural frequency at the left border of the interval and the oscillator with the minimal negative natural frequency at the right border of the interval. The process of network desynchronization begins with them. To illustrate the desynchronization process, we took the value of the parameter $\beta$ outside the synchronization interval, but close to the left border: $\beta=4.15$. It turned out that for these value of the parameter $\beta$, the oscillators are firstly synchronized, but then the oscillators one by one begin to separate from the synchronous group, starting with the oscillator with the highest natural frequency and then descending the natural frequency. This process is illustrated in FIG. \ref{fig:time_freqs}a, where the frequency of the highest-frequency oscillator, the frequency of the 25th oscillator, the common frequency and the order parameter $R_2$ are shown in time. When the first oscillator separates from the synchronous group, the order parameter $R_2$ decreases. Because of this, according to the formula \eqref{left_border_3D}, the stability thresholds for all other oscillators are shifted, and the second oscillator crosses the boundary of the synchronization interval. Because of this, it separates too from the synchronous group. And then, in turn, all the oscillators are separated from the synchronous group due to the fact that their thresholds are shifted in time (FIG. \ref{fig:time_threshold}a). Even though some of them temporarily rejoin the synchronous group, the process of desynchronization has already started and eventually leads to the complete disintegration of the synchronous group. Thus, in system \eqref{sys_3D}, when the stability of the equilibrium state for one oscillator is lost, an abrupt loss of stability occurs for the entire network.

For comparison, FIG. \ref{fig:time_freqs}b and FIG. \ref{fig:time_threshold}b show that in the case of pairwise couplings, the separation of one oscillator from a synchronous group is not enough to cause the entire network to lose stability, and there remain oscillators whose dynamic threshold $\beta_{left}$ never crosses the boundary of the synchronization region.

\section{\label{sec:conclusion} Conclusion}

Firstly, a class of adaptation functions is found for which a synchronous oscillation mode exists in the network of phase oscillators with triadic couplings. These functions are concentrated in a fairly wide range of adaptation parameter and include Hebbian plasticity. Secondly, we have shown that the destruction of the synchronous mode occurs differently for networks with pairwise couplings and with higher-order interactions. In the first case, a chimera state is realized. In the second case, the destruction of the synchronous state occurs more abruptly, and the chimera state is not formed. Thirdly, the patterns of formation of synchronization and desynchronization modes are determined. We have found that the desynchronization process starts with the highest frequency oscillators and then proceeds to lower frequencies.

It is worth noting that we also took other initial conditions, such as $\theta_i = \frac {\pi} {2} - \beta$ and $\theta_i = \frac {3\pi} {2} - \beta$. This led to the same results. This means that the results are robust to a change in the initial conditions.

\begin{acknowledgments}
This work was supported by the Russian Science Foundation under Project No. 23-42-00038.
\end{acknowledgments}

\section*{DATA AVAILABILITY}
The data that supports the findings of this study are available within the article.

\nocite{*}
\bibliography{aipsamp8}

\end{document}